# Concentration-control in all-solution processed semiconducting polymer doping and high conductivity performances

*Khaoula Ferchichi[1,2], Ramzi Bourguiga[2], Kamal Lmimouni[1], Sébastien Pecqueur[1,*]*

1. Univ. Lille, CNRS, Centrale Lille, Yncréa ISEN, Univ. Polytechnique Hauts-de-France, UMR 8520 - IEMN, F-59000 Lille, France.
2. Laboratoire physique des matériaux, Faculté des Sciences de Bizerte, Université de Carthage, 7021 Jarzouna–Bizerte, Tunisia

\* sebastien.pecqueur@iemn.fr

**Abstract**

Simultaneously optimizing performances, processability and fabrication cost of organic electronic materials is the continual source of compromise hindering the development of disruptive applications. In this work, we identified a strategy to achieve record conductivity values of one of the most benchmarked semiconducting polymers by doping with an entirely solution-processed, water-free and cost-effective technique. High electrical conductivity for poly(3-hexylthiophene) up to 21 S/cm has been achieved, using a commercially available electron acceptor as both a Lewis acid and an oxidizing agent. While we managed water-free solution-processing a three-time higher conductivity for P3HT with a very affordable/available chemical, near-field microscopy reveals the existence of concentration-dependent higher-conductivity micro-domains for which furthermore process optimization might access to even higher performances. In the perpetual quest of reaching higher performances for organic electronics, this work shall greatly unlock applications maturation requiring higher-scale processability and lower fabrication costs concomitant of higher performances and new functionalities, in the current context where understanding the doping mechanism of such class of materials remains of the greatest interest.

**Introduction**

In today's state-of-the-art of semiconductor technologies, organic semiconductors have both demonstrated their ability to achieve technological applications at the consumer-electronic market-level [1,2], and continues to inspire the scientific community by unlocking truly disruptive abilities to realize next-generation electronics[3–5]. For their potentials for low-cost electronics[6], soft processing, and device added features such as flexibility or transparency[7], these materials have been widely investigated in a range of microelectronic and optoelectronic architectures such as field-effect transistors[8], light-emitting PN-diodes (OLEDs)[9], Schottky diodes[10] and single carrier sensing devices[11]. In any of these applications, significant improvements in the electrical performances[12–14] can be achieved by means of molecular doping (i.e., chemical doping by incorporating highly electron-accepting/donating molecules into the organic semiconductor). While doping is one the key-enabling technology behind conventional CMOS electronics (stands for Complementary Metal-Oxide Semiconductor), it remains superficially exploited in some of the organic semiconducting applications [14], while it can greatly optimize both charge-carrier injection and transport in semiconductor devices diminishing the overall device power consumption.

Although most polymers are electrical insulators, exceptions show outstanding electrical performances, with only two orders of magnitude less conductive than metals, while they maintain unique properties required for plastic- or transparent-based electronics. As such an example, poly(3,4-ethylenedioxythiophene)-poly(styrenesulfonate) (PEDOT:PSS) has been extensively investigated as a flexible and transparent p-type material with conductivities up to 1700 S/cm for PEDOT:PSS[15]. But similar to other common conducting polymers such as polypyrrole (PPy, up to 90 S/cm)[16] or polyaniline (PANI, up to 24.39 S/cm)[17], their ionic nature responsible for their high conductivity implies their use in water as a solvent, which threatens the performances of the devices for optoelectronics applications. Other aspects of these materials are hindering their use in optoelectronics to the broadest range (which imply the formation of high yield of excitons with specific energies), such as depth of the HOMO level or light-absorption coefficient, for which these p-type materials do not meet the strong requirements to be used for other than interfacial hole-injecting layers[18,19]. Meeting all the necessary properties for an optimized organic electronic devices in injection, transporting, blocking of holes and electrons and/or exciton formation/recombination frequently requires the conjunction of multiple specific semiconductors in a sophistically engineered stack, where the conductivity of most of these layers in the stack have to be increase to lower the overall device resistance and its power consumption[20–22].

Among all organic semiconductors, poly-3-hexylthiophene (P3HT) is a benchmark polymer semiconductor material, which has been widely studied in many device technologies such as photovoltaic generators[23,24], photo-sensing diodes[25], organic light-emitting diodes[26,27], p-type transistors[28,29], thermoelectric generators[30,31] and more recently synaptic elements[32,33]. Therefore, being able to dope materials such as P3HT is essential for having a control on the performances of all these organic electronic applications. In particular, much interest has been devoted on using tetrafluoro-tetracyano-quinodimethane (F4-TCNQ) as a p-dopant for P3HT among other semiconductors, for both the solubility in organic solvents and the high electron affinity (5.2 eV) of this small molecule[34,35].

To achieve the best uniformity in the doping profile within a doped organic semiconductor thin-film, a first strategy is to blend both materials in a solution prior its deposition on a substrate. While this strategy allows the control of the doping concentration necessary to regulate the electronic performances of the doped semiconductor, it requires for these materials to observe high solubility in a common solvent. In the case of P3HT doping by F4-TCNQ, this method leads to material aggregation in both the solution [36,37] and the dried thin film [38,39] for high doping concentrations, which results in a much lower doping yield.

To the best of our knowledge, the highest electrical conductivity achieved by such "solution mixed process" is 8 S/cm [38]. Recent efforts are focused on going beyond this practical limit: To achieve higher doping yield for a specific polymer semiconductor such as P3HT, two strategies are adopted: either investigate on unconventional processes (to optimize the integration of F4-TCNQ into P3HT thin-film) or on new electron-acceptor dopants (showing better processing capabilities to p-dope P3HT).

By sequential doping, in which the polymer is first deposited prior the dopant with an orthogonal solvent, electrical conductivities higher than those with solution-mixed process have been achieved[38,40,41]. Recently, it has been shown that sequential doping can be performed from F4-TCNQ deposited from the vapor phase[42–46], higher electrical conductivity up to 48 S/cm has been reached [45]. Lim and co-workers [45] have achieved this record by estimating the amount of dopant by the time on which the film was exposed to the vaporized dopant, from optical absorption spectra, the high concentration that leads to this value is 9.4±0.1% S/cm. Furthermore, Hu and co-workers[47] showed that electrical conductivity in P3HT doped with F4-TCNQ can be enhanced by means of one dimensional nanomaterial templates. The blended solution of P3HT and F4-TCNQ is dispersed on an anodized alumina template, and then annealed at 250°C under vacuum for six hours. The molten P3HT/F4-TCNQ is then introduced into the template by capillary force. The sample was then cooled gradually to solidify the P3HT/F4-TCNQ composite and form 1D nanowire. Sakr and co-workers[48] achieved a high electrical conductivity of 22 S/cm, by combining the controlled orientation/crystallization of polymer films by high temperature rubbing with the soft doping method by spin coating a solution of acetonitrile-F4-TCNQ on top of the oriented polymer films.

While all these techniques have enabled reaching conductivity values which are one-fold higher, the lack of control in the doping concentrations threaten development perspectives.

To preserve the control of the doping rate, alternative materials to F4-TCNQ are also investigated to dope P3HT[49–51]. Results obtained with 30% $FeCl_3$ doping show higher conductivities than the current state-of-the-art F4-TCNQ:P3HT systems doped by blending process, up to 15 S/cm[50]. High electron-affinity molybdenum-based organometallic electron acceptors such as Mo(tfd)$_3$[52], or Mo(tfd-CO2Me)$_3$ have also shown doping on polymers and small molecules[50,53], with conductivity respectively up to 5 and 2 S/cm in P3HT[50]. Low-cost polyoxometalates have also been used in the literature such as phosphomolybdic acid or phosphotungstic acid to dope P3HT with conductivity up to 7±0.5 S/cm [54] by post immersion of P3HT film into a polyoxometalate solution in nitromethane. Recent reports focused on the electron-acceptor chemical doping of P3HT using the strong Lewis acid tris(pentafluorophenyl)borane (BCF)[49]. However, the conductivity is still low about $10^{-3}$ S/cm with $10^{-2}$ dopants per P3HT monomer. These recent results translate a persistent motivation in finding new doping alternatives for polymer semiconductors to optimize the compromise between affordability, processability and performances.

In this study, we characterize the doping of P3HT by an organometallic electron-acceptor which is extensively used as catalyst for its strong Lewis acidity: Copper(II) trifluoromethanesulfonate, (Cu(OTf)$_2$). Solution blends of P3HT and Cu(OTf)$_2$ has been prepared, and then deposited by spin coating, thin-films with high molar concentration ratio up to 0.33 (one dopant per two monomers units) has been prepared through this process. Moreover, this material being an air stable, soluble in common organic solvent as well as P3HT, commercially affordable at a 10€/g scale, it shows remarkable electronic and optical properties as a p-dopant suitable for various device applications. We characterized the electrical and optical properties of the new materials doped at the several molar concentrations and characterized the materials morphology to demonstrate the mechanism underlying the high electrical conductivity up to 21 S/cm: six times higher than literature for P3HT doped via a concentration-controlled process.

**Experimental**

For the whole experience, regioregular poly(3-hexylthiophene) (rr-P3HT - purity, 99.995%) and copper(II) trifluoromethanesulfonate (Cu(OTf)$_2$ - purity, 98%) were purchased from Sigma Aldrich and used as received without further purification. Cu(OTf)$_2$ was used and kept under inert atmosphere.

P3HT has been dissolved in chlorobenzene and solubilized as an orange solution (concentration of 10 mg/ml). Cu(OTf)$_2$ has been formulated from a mixture of dry solvents chlorobenzene:isopropropanol (7.2:1 v/v) and results in a pale blue solution. The solvent ratio has been optimized to avoid gelation after the addition of the dopant solution into the P3HT one. Isopropanol was used to promote the solubility of Cu(OTf)$_2$ as protic solvent favoring ion dissociation and lone-pair donor coordinating the Lewis acid **[55]**, while excess of isopropanol promotes P3HT's gelation. For the preparation of the blend solutions, appropriate amount of prepared Cu(OTf)$_2$ solution was added to the prepared solutions of P3HT, to achieve the desired dopant molar fraction $X_{mol}$ ($X_{mol}=n_{dopant}/(n_{dopant}+n_{SC})$). The addition of the dopant to the polymer solution led to a blue color solution with no precipitate, which allowed us to use convenient solution processing by spin coating. Fresh solutions of pristine and doped P3HT were filtered with a 0.45 µm PTFE filter, and then deposited by spin coating on a Si/SiO$_2$ substrate with patterned electrodes, using a two-step spin condition of 500 rpm for 60 s followed by 1000 rpm for 30 s. The fabricated thin films were then backed at 100°C in vacuum for one hour to remove any residual solvent. An average thickness of 42±5 nm for the films doped at different molar fraction were evaluated by profilometry, with ± 20% of relative uncertainty for each measured layer.

We studied the effects of Cu(OTf)$_2$ on the electrical conductivity of the P3HT polymer first based on current voltage measurements of 2-terminal interdigitated gold structures. Electrical measurements in an inert environment, with an Agilent 4156 parameter analyzer. Resistance has been first measured by transmission line method (TLM) using the different device geometries (Figure S1), with inter-electrode spacing (L) ranging from 1 to 50 µm and width W = 1 mm. Absorption spectra for spin-coated polymer thin films doped at different molar ratios were recorded with a Lambda 800 UV−visible absorption spectrometer on referenced glass substrates in air and at room temperature. Conductive atomic force microscopy (C-AFM), was performed in air with the peak force TUNA module (ICON, Bruker). The probe used for the experiments is a silicon probe coated Pt/Ir (Platinum/Iridium) with a radius of 25 nm, and with a spring constant of 0.4 N/m. Peak force tapping mode modulates the z-piezo at 1 kHz with a peak force amplitude of 100 nm. The bias voltage applied between the probe and the sample is 1 V.

**Results and discussions**

Cu(OTf)$_2$ has already been used as a dopant for both conducting polymers from solution-processing **[56]**, and for small-molecule organic semiconductors by vacuum co-evaporation, with a remarkable temperature stability up to its sublimation point at 360°C under $10^{-6}$-$10^{-5}$ mbar pressure range **[57]**. Here to study its influence on P3HT, the conductivity of the doped layer has been systematically extracted from the current voltage electrical measurement performed on the different concentration of dopant:semiconductor systems.

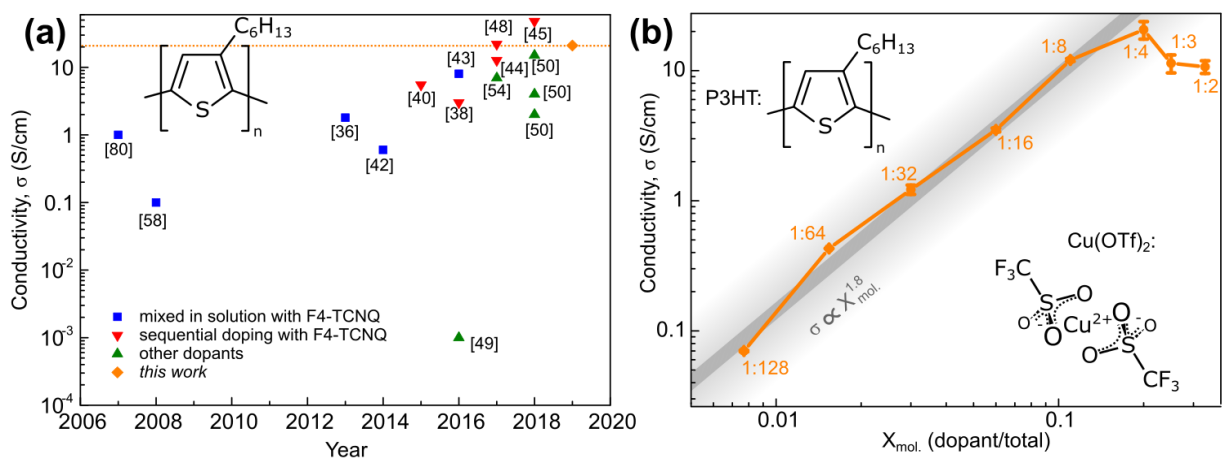

**Figure 1.** (a) Recent state-of-the-art of P3HT p-doping, (b) Conductivity dependence on the dopant molar fraction.

*Figure 1* shows the electrical conductivity increase with the dopant molar fraction for dopant molar fraction ranging from 0.0077 to 0.33 (corresponds to dopant:monomer-unit ratios from 1:128 to 1:2). The pristine P3HT polymer showed a conductivity value of about 1 mS/cm, which is one to two orders of magnitude higher than previously reported values[36,58]. The relatively high conductivity value for the pristine material has been attributed to $O_2$-doping, giving the fact the P3HT material has been pre-handled in air in addition of the assessed effect of molecular oxygen on P3HT among many other hole transport materials. P3HT's conductivity increases superlinearly with the doping at low concentration, according to a power law of exponent n = 1.8 (*Figure 1b*). As low-concentration doping usually promotes a linear increase of the conductivity with the concentration[59], the superlinearity observed with the doping concentration is attributed to the doping-level dependent mobility of P3HT in addition of the carrier density increase: As shown by Arkhipov and coworkers, P3HT's hole mobility can increase by a power up to three with the doping concentration thanks to HOMO profile alteration by the dopant due to Coulomb well overlapping[60]. P3HT's conductivity doped with $Cu(OTf)_2$ shows a maximum of 20.6±3.2 S/cm for a molar ratio of 0.2 between dopant and monomer-unit. In contrast, higher doping ratios (> 0.2) lead to a noticeable decrease of the conductivity toward a limit of 11 S/cm, about twice lower than the maximum.

This lowering of the P3HT conductivity at such concentrations of dopant is similar to previous reports of molecular doping for different types of organic semiconductors[34,36,61,62]. In the case of F4-TCNQ-doped P3HT, it has been reported that the doping concentration can limit the maximum achievable conductivity[36,63]. For instance, Pingel and coworkers simulated that F4-TCNQ molecules can interact with multiple thiophene units of a same chain to form a single complex with only partial charge-transfer rate[63]. Also, increasing the doping concentration to a value higher than this limit can result in the formation of aggregates in the case of F4-TCNQ[37,63,64]. Such similar phase segregation between P3HT with the aggregating $Cu(OTf)_2$ could potentially explain the observed lower conductivity.

**Table 1.** Summary of dopant concentration at maximum conductivity reported in the literature.

| Doping method | Concentration (mol%) | Conductivity (S/cm) | Reference |
|---|---|---|---|
| Mixed solution | 17 | 8 | [38] |
| Sequential from CB | 8 | 3 | [38] |
| Sequential from $CH_3CN$ | 5 | 3 | [38] |
| Vapor doping | 10 | 48 | [45] |
| Mixed solution | 25 | 20.63 | **This work** |

Table 1 reports on the highest electrical conductivity achieved in the literature for F4-TCNQ doped P3HT, and the molar percentage ($\%^{mol}$) that allows to achieve this value. With unconventional process such as vapor doping, the highest value achieved in the literature is 48 S/cm for 10% concentration (P3HT:F4-TCNQ). From mixed solution the highest value is about 8 S/cm for 17%. In our case, we demonstrate experimentally the feasibility to enhance

P3HT's conductivity by 20.6 S/cm via a concentration-controlled process, allowing the control of the layers electrical performance via predefined concentration as well as ensuring the dopant's concentration homogeneity within the layer, essential to control PN diodes performances for instance.

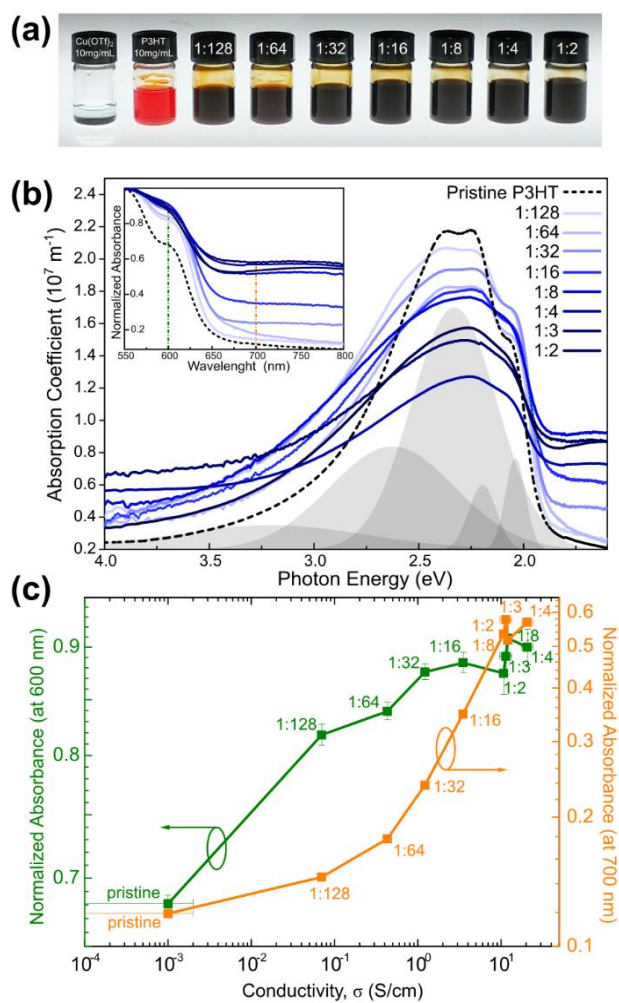

**Figure 2.** (a) Picture of the different doped solutions used in the experiments, for which each of them shows strong light absorption in the visible range compared to the absorption of the pure dopant and pure P3HT mother solutions. (b) UV-vis absorption spectra of P3HT thin-films, doped with $Cu(OTf)_2$ at different molar ratio (the distributions are Gaussian fittings from the pristine P3HT data – see supplementary information Figure S2 for details). (c) Relative intensity versus conductivity at the band-gap absorption shoulder (600 nm) and below the band-gap energy (700 nm).

To gain further insight on the mechanism underlying in the conductivity increase and gather information on the interaction between $Cu(OTf)_2$ and P3HT, UV-vis absorption spectroscopy has been carried out on the doped P3HT films. Experimentally, the $Cu(OTf)_2$-doping of P3HT is characterized, in both the solution and the dried thin-film, by a pronounced color-change from orange to blue (*Figure 2a*), gradually with the increase in Cu(OTf)$_2$ concentration. *Figure 2b* shows the optical absorption coefficient (as measured absorbance over evaluated layer thickness) of undoped and Cu(OTf)$_2$-doped P3HT layers. This spectra reveal two distinctive trends with the dopant concentration increase which are observed for other doped conducting polymer systems in the literature **[38,40,54,58,65]**: A gradual decrease of the pristine-P3HT absorptions with doping, and an increase of absorbance in the sub-gap region of the spectrum. Although both trends are observed on the absorption coefficient for the nine different variations in doping concentrations, punctual deviations from monotonicity are observed. Origins from these deviations are hypothetically of different kinds: sample-dependent paraxial deviation, layer-dependent thickness deviation due roughness and yield of charge-transfer variation affected by a concentration-dependent dopant percolation. Considering the averaged thickness of 42 nm for the thin-films, deviations to the normality of the substrate might origin to interference causing systemic variations of the measured absorbance. Also, the uncertainty of the layer thickness associated to limitation of the profiler and the deposition process induces comparable relative deviations for the extinction coefficient. Shall also be pointed that light scattering due to the surface roughness might also affect the trends with the doping concentration (visually evidenced experimentally, with a more pronounced effect on the samples with a higher doping concentration).

To circumvent most of these substrate-specific effects, we compared the optical data by the mean of the normalized absorbance as the ratio of two measured absorbance taken from the same sample in the same experimental run (displayed as an inset in *Figure 2b*). It also allows referencing each sample to its doping-concentration specific semiconductor density, for which the diminishing of the P3HT band-gap absorption is non-negligibly influenced by the dilution of the P3HT by Cu(OTf)$_2$ at the highest concentrations. On the normalized absorbance, we observed a clear correlation between increased relative intensity at both the lowest energy absorption (700 nm) and the one below the band-gap (600 nm) with the electrical conductivity (*Figure 2*c). Since the relative absorption at 700 nm simultaneously compares the increase of the sub-gap region and the decrease of the major absorption peak as the ratio of the one on the other, its value is a quantitative doping indicator as observed from the relative-absorption/conductivity correlation. Despite the deviations with the doping concentration which were previously mentioned, this monotonic behavior with the conductivity has also been observed with the measured absorption coefficient (*Figure S3*). In both absorption/conductivity correlations (*Figure 2*c), we observed that both properties converge to a limit at high dopant concentrations: As the conductivity of the film reaches its maximum for $X_{mol.} > 0.2$, so does the relative absorption at 600 and 700 nm. This correlation suggests that before the conductance limitation with the doping, increase in conductivity is associated to the electronic transitions observed in the absorption spectra and not only to the mobility enhancement with the doping concentration (Arkhipov and coworkers)[60]. But for concentration $X_{mol.} > 0.2$, the decrease in conductivity cannot be associated to a lowering of carrier density, since the relative absorptions do not decrease with the dopant concentration. This last suggests that the decrease in conductivity observed at high concentration is rather promoted by a mechanism which affects the material mobility (transport property) than the carrier density (characterized by absorption spectroscopy).

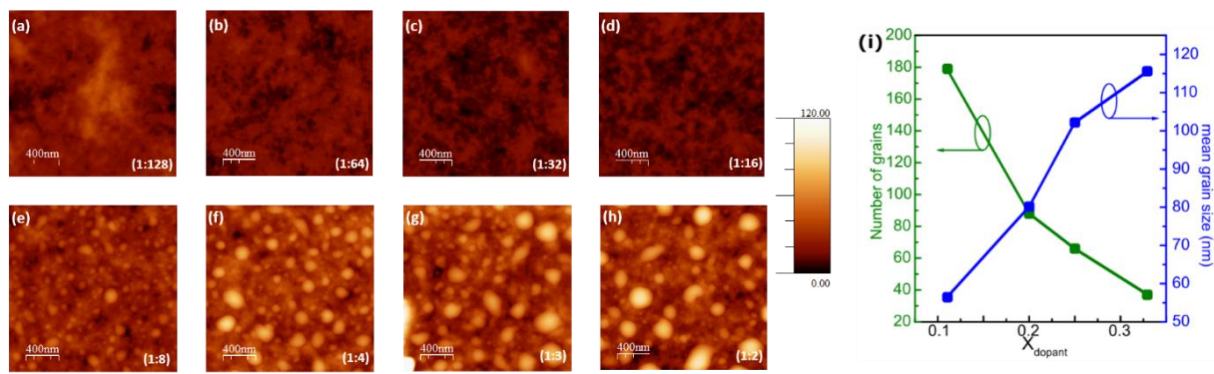

**Figure 3.** AFM topography images (2x2 µm²) of a)-h) P3HT doped with different molecular fractions, ranging from (1:128) to (1:2). (i) Molar ratio dependency on the number of grains and mean grain size at high Cu(OTf)$_2$ doping (e-h).

To verify it, atomic force microscopy (AFM) has shown various film morphology for the different doped layers (*Figure 3*). At low dopant concentrations (*Figure 3 (a)-(d)*), films appear to be uniform at the micrometer scale, with no apparent aggregation and a surface roughness of root mean square (RMS) below 2.5 nm (about 5% of the nominal thickness value evaluated by profilometry). At high concentrations (*Figure 3 (e)-(h)*), P3HT films show the formation of sub-micrometric aggregates which increase in size with the doping concentration, with a RMS roughness which is two-to-three times higher, up to 6.5 nm for 1:2 doping ratio. One can also observe the non-isotropy of these aggregates by their shape and their flatness (their height are statistically smaller than their planar characteristic dimension). This change in surface roughness with the doping concentration also validates the earlier hypothesis on the morphology of the layers affecting the absorption coefficient due to light scattering. Moreover, it should be pointed out that although the RMS of the layers is lower or comparable to the thickness incertitude evaluated by profilometry, we can observe from *Figure 3 (a-h)* that some aggregates can achieve height which are substantially higher than the 42±5 nm thickness evaluated by profilometry, suggesting that the incertitude of the thickness value might be further source of absorption coefficient deviation. This noticeable increase of the surface roughness with the doping concentration provides a direct evidence for the Cu(OTf)$_2$–doped P3HT morphology to limit its electrical performances at the material level rather than at the molecular scale. Similar behaviors have also been reported in the literature for several doped organic semiconductor like for F4-TCNQ doped P3HT[38,66,67] and PCPDTBT[67], and also in the case of organometallic dopant[68]. As the density of dopant increases in the polymer layer, we observed an increase of the aggregation rate of these nuclei, by their decrease in number and increase in average size (*Figure 3i*). Over the four samples from 1:8 to 1:2 doping ratios, trends from the increase in size and decrease in number is monotonic, from 45 aggregates/µm² with a mean size of 56 nm at 1:8 molar ratio, to 10 aggregates/µm² with a mean size of 115 nm at 1:2 molar ratio.

The correlation between aggregation rate with concentration of dopant clearly evidences a concentration-dependent phase segregation between two materials, resulting from the addition of dopant in the polymer. Hypothetic explanations on the nature of the segregated phases could be: (i) a separation between pure semiconductor and doped semiconductor, as reported in previous studies[34,36,69], or (ii) a separation between semiconductor and dopant domains, forming respectively doped and undoped domains in the thin-film[34,36,38].

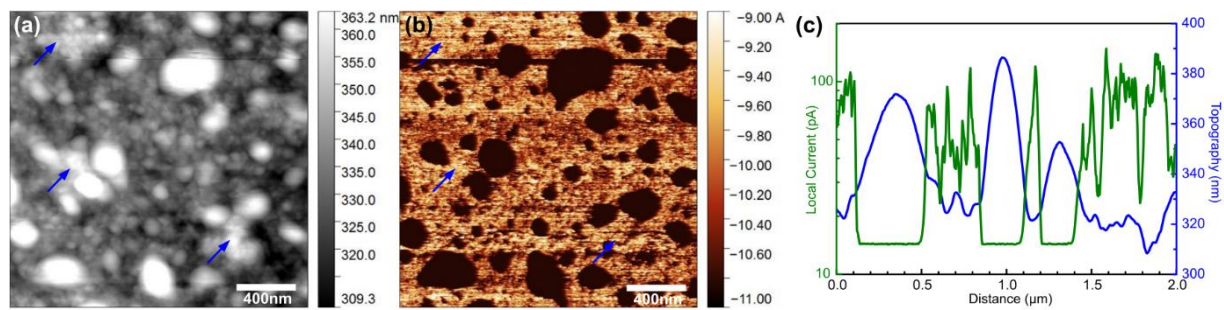

**Figure 4.** (a) AFM topography of the (1:2) $Cu(OTf)_2$-doped doped P3HT (1:2), (b) Conducting AFM, (c) Local current and topography versus distance.

To identify the electronic nature of the different micrometric phases of the doped semiconductor composite, $Cu(OTf)_2$-doped P3HT layers has been probed by conducting atomic force microscopy (C-AFM). It appeared from the (1:2)-$Cu(OTf)_2$-doped P3HT layer (*Figure 4*) that the material's topology rules the electrical property of the material at the sub-micrometer scale, such that the aggregates (assumed to be dopant-rich because of the concentration-dependent size) show to be at least one order of magnitude less conducting than the dispersing phase (assumed to be semiconductor rich). This correlation between conductivity and topology at the sub-micrometer scale confirms the actual doping of Cu(OTf)$_2$ on the P3HT: As the dopant-rich aggregates are more insulating than the semiconductor-rich phase, it is legitimate to assume the high conductivity of the composite material to be due to the actual conductivity enhancement of the P3HT conducting-polymer phase, rather than the adjunction of Cu(OTf)$_2$ hypothetically conducting in its pure form (note that thermally evaporated transition metal oxides such as MoO$_3$ can be both p-dopant and exhibit metallic conductivity due to decomposition)[70]. In completion with the doping yield diminishing due to dopant aggregation, the insulating nature of the dopant-rich aggregates shall also have a negative impact on the charge carrier mobility by the scattering of holes at the micrometer scale due to the presence of dense non-conducting domains. We also identified domains on the C-AFM images which showed aggregate topologies with high conductance (blue arrows on *Figure.4a-b*) which opens on the existence of another phases, though apparently minor compared to the first two identified isolating "dopant-rich" and conducting "polymer-rich" phases.

**On Cu(OTf)$_2$'s doping mechanism(s)**

Lewis acids are definitely a class of electron accepting chemicals which demonstrates as much promises for p-doping organic semiconductors[49,71,72], as interests for identifying the driving force promoting their electronic transfer with organic semiconductors[73]. Still remaining an intriguing research topic[73], their lone-pair accepting nature is incompatible with the single-electron charge transfer mechanism where the charge carrier is only bound electrostatically. The ambivalent nature of Cu(OTf)$_2$ to be both a redox-active copper(II) salt (triflates as weakly-coordinating anions) and a Lewis acids (which forms coordination compounds with poor donors such as benzene of toluene) opens many possibilities to justify the physical nature of the electron-acceptor doping mechanism (*Figure 5*).

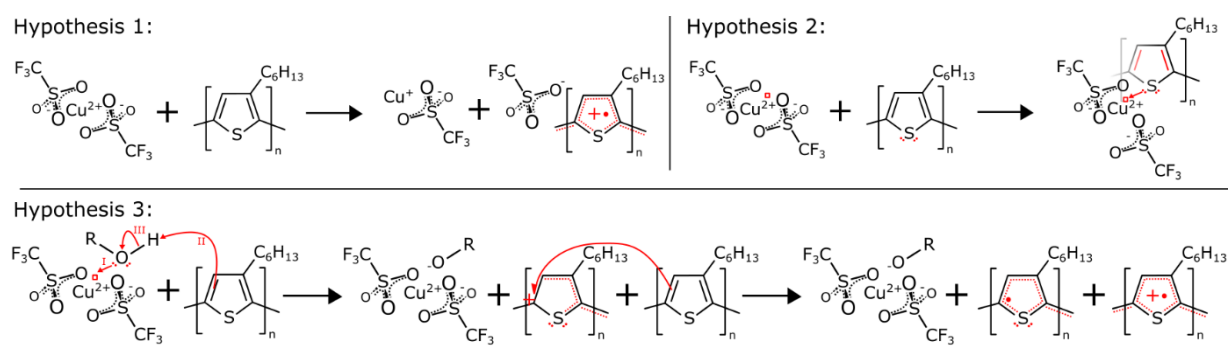

**Figure 5.** Three proposed mechanistic hypothesis based on Cu(OTf)$_2$'s and P3HT's properties and in light of the current state of the art on molecular doping.

A first hypothesis lies on the dopant dissociation promoting integer charge transfer (*Figure 5 – Hypothesis 1*): As a source of triflates, the solution processing can promote ion dissociation of Cu$^{2+}$$_{(solution)}$ with OTf$^-$$_{(solution)}$ to undergo a redox-driven mechanism between solubilized copper(II) ions and solubilized semiconductor molecules, similarly as the doping mechanism of Fe(OTf)$_3$ on PEDOT[74,75]. As for PEDOT[75], Cu(OTf)$_2$-doped polythiophene and polypyrrole can trap copper-free OTf- ligands after washing of the organic phase (0.2 and 0.3 mole of OTf$^-$ per monomer unit)[56]. This favors the hypothesis that the continuous polymer-rich phase that has been identified in *Figure 3* (for different concentrations) and *Figure 4* (to be particularly conductive) contains free triflates responsible for its high conductivity, while the aggregates would be mainly constituted of a lower oxidation degree Cu(OTf)$_{2-x}$. Also, given that such triflate/polythiophene can form crystals, it opens the possibility to observe also high-conductivity aggregates (blue arrows *Figure 4*).

A second hypothesis lies on the promotion of Lewis adducts (*Figure 5 – Hypothesis 2*): As the Cu(OTf)$_2$-doping occurs also at the lowest concentrations (*Figure 1b*) for which no phase segregation has been evidenced by C-AFM (*Figure 4*), one can question on the necessity of lower-oxidation-degree dopants observed as aggregates. Moreover, Cu(OTf)$_2$ has already shown p-doping when co-evaporated with high phase-transition temperature arylamines under $10^{-6}$ mbar[57,76]. In such conditions where Cu(OTf)$_2$ does not benefit from polar solvents to dissociate its ions, the formation of Lewis acid-base adducts between the semiconductor with the whole Cu(OTf)$_2$ dopant has been proposed to explain the obtained $5 \cdot 10^{-4}$ S/cm conductivity recorded under inert atmosphere[57]. In these co-evaporated systems, the layers' relative absorption below the semiconductor bandgap displays the same trend with Cu(OTf)$_2$ concentration as in our case for solution-processed P3HT (*Figure 2*). This suggests that the electronic transitions, which are characteristic of the semiconductor doping, does not depend on the process whether ion-dissociating solvents are used or not. Such wide absorptions at low energies are very characteristic of hybrid charge-transfers and distinct from radical-ion absorption for integral-transfers [14], where the dopants hybridize with the semiconductor to feature a new narrow bandgap system derived from the molecular orbital theory of Lewis acid-base adducts [77–79]. On the question of whether such mechanism can promote sufficient charge dissociation by thermal activation to significantly dope the semiconductor, we showed that the Cu(OTf)$_2$ Lewis acids can achieve higher conductivity than solution-processed F4TCNQ, although F4TCNQ is known to undergo integral charge-transfer with P3HT (and hybrid only on short oligothiophenes)[14,80]. We also note that, in our case with Cu(OTf)$_2$, the concentration optimum is more characteristic of integer charge-transfers than the 50% optimum of hybrid ones[14], but we evidenced in our case that morphology might have hindered doping yield furthermore by the formation of insulating aggregates.

As a third hypothesis (*Figure 5* – Hypothesis 3), participation of a third-party element promoting integer-charge transfer on the whole dopant without dopant ionization: Recently in the case of BCF-doping on thiophene-containing semiconductors, it has been theoretically and experimentally evidenced that the strong electrophilicity of the boron-containing Lewis acid can promote dissociation of water to protonate the conducting poly/oligomers, which in a second non-limiting step promotes an electron transfer from a neutral polymer chain to the protonated one[73]. More interestingly, the chemical modification of the thiophene-containing semiconductors by substitution of a vicinal aromatic carbon by nitrogen disables the low absorption energy and decreases the layers' conductivity [73]. Our experimental conditions favor also this hypothesis by the presence of isopropanol to ease the solubility, and the use of P3HT as a poor Lewis base. Based on the same assumption that protonated P3HT is less stable than the radical, doping is indirectly promoted by the proton transfer from the alcohol for which the alcoholate is further stabilized by the highly electrophilic copper(II) core. The absorption trends of $Cu(OTf)_2$:P3HT layers are again very comparable to the one of BCF-doping in the case of a poorly Lewis-basic semiconductor (*Figure 2*). Also, as the solvent promotes integer-charge dissociation, phase separation between a dopant-rich domain and a polymer-rich one can occur (*Figure 4* and *Figure 5*). Shall be pointed out that $Cu(OTf)_2$ doping of nitrogen containing semiconductors has been reported[57,76], which shows that the use of $Cu(OTf)_2$ is not strictly limited to thiophene-containing aromatics such as P3HT and can be applied to semiconductors which have mild Lewis basicity.

In light of the morphological, electrical and optical data, and as these three mechanisms are not antagonist and can be concurrent to favor the electron charge-transfer with the polymer, it seems highly probable that the ambivalent Lewis acidic / redox active $Cu(OTf)_2$ salt exploits many of its aspects to p-dope up to record conductivities for fully solution processed doped-P3HT layers.

**Conclusion**

In this work, record conductivity values for P3HT have been achieved via an entirely concentration-controlled solution-process, by the usage of an organomettalic p-type dopant $Cu(OTf)_2$. Electrical and optical properties of the layer confirmed the p-doping effect by the conductivity-dependent low-energy absorption characteristic of Lewis-acid semiconductor p-doping. Atomic-force morphological studies informed us furthermore on the nature of the new p-doped P3HT material, by the formation of dopant-concentration dependent insulating aggregates: while it promises encouraging possibilities to increase furthermore P3HT's conductivity by tuning the nano-domains morphology, it confirms also Cu(OTf)$_2$'s genuine doping by charge-transfer to the semiconductor. By confronting Cu(OTf)$_2$:P3HT properties to current state-of-the-art mechanisms on organic semiconductor doping via Lewis acids, we propose that because of the ambivalence of Cu(OTf)$_2$ as both a redox active salt and as a strong Lewis-acid, the dopant exploits multiple processes to favor the donor/acceptor charge-transfer, usually yielded at different levels by mechanisms such as frontier-orbital overlapping at the molecular scale or phase segregation at the micro-scale. Reaching high conductivity performances on one of the most extensively used conducting polymer, this finding opens new perspectives for upcoming higher-performances and also for deeper understanding for this new class of ambivalent dopants for organic semiconductors.

**Supporting Information Description**

In the supporting information, we give the plots of the resistance versus the electrodes spacing length L, for the different dopant molar ratios $X_{dopant}$, the fitting of the pristine P3HT's absorption spectrum and the absorption coefficient versus conductivity below the band-gap energy.

**Acknowledgements**

This work was supported by European Project Interreg Luminoptex and ANR context project. We thank the French National Nanofabrication Network RENATECH for financial support of the IEMN cleanroom.

**Table of contents images**

# Supporting Information

Figure S1 plots the resistance versus the electrodes spacing length L, for the different dopant molar ratios $X_{dopant}$.

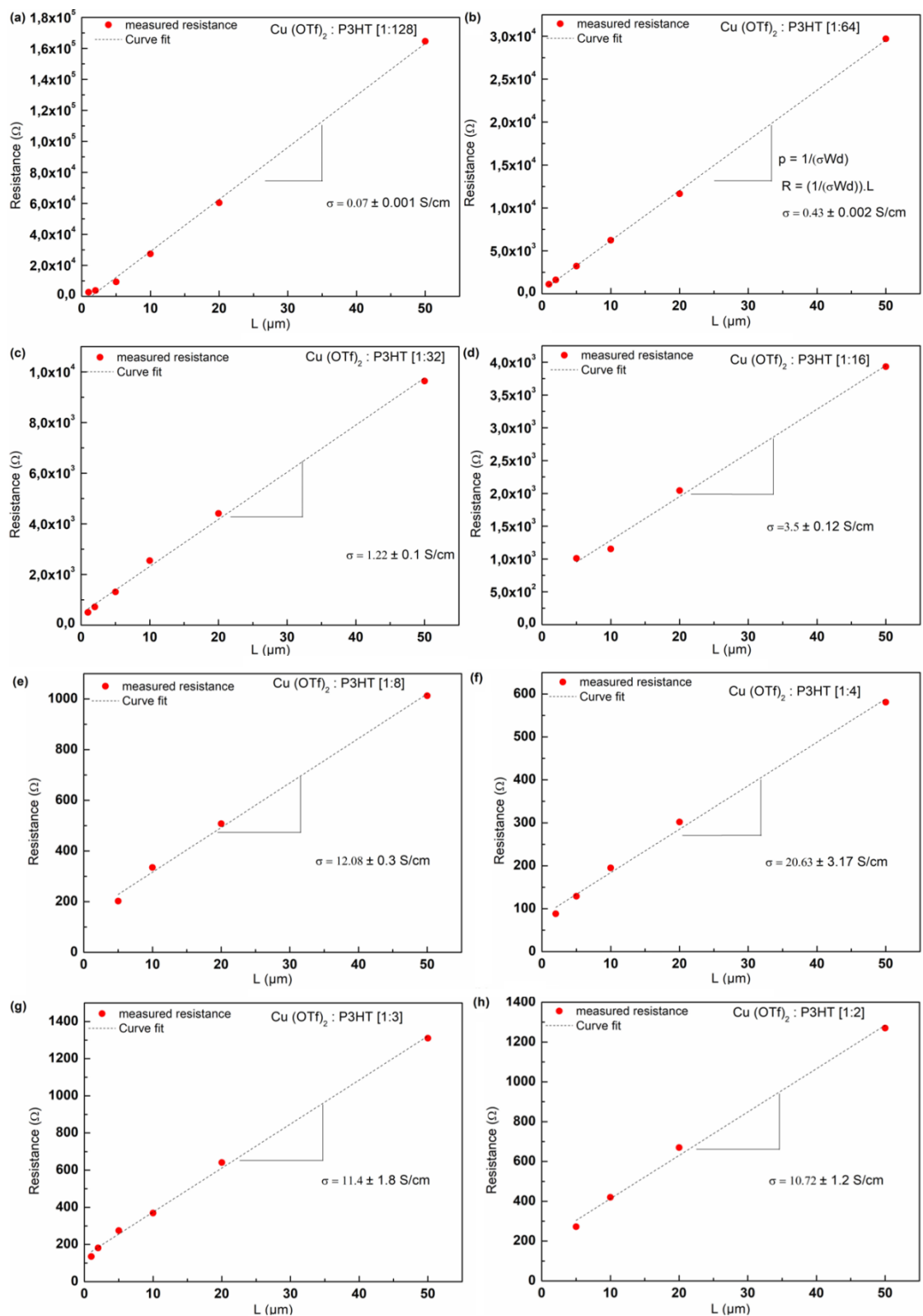

Figure S1. Resistance plot versus length

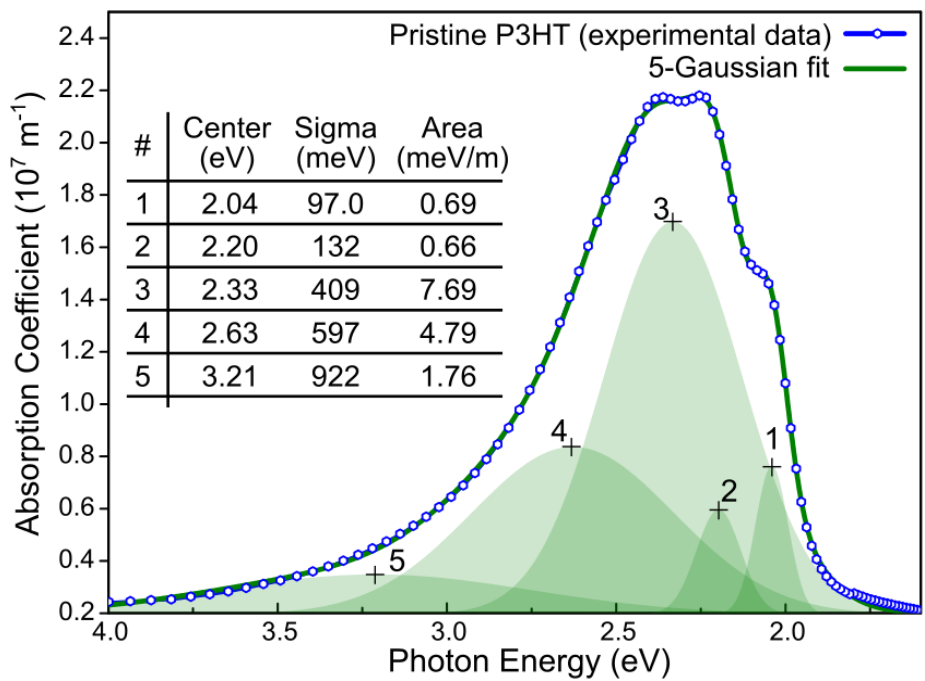

Figure S2. Fitting of the pristine P3HT's absorption spectrum with 5 Gaussian density distributions

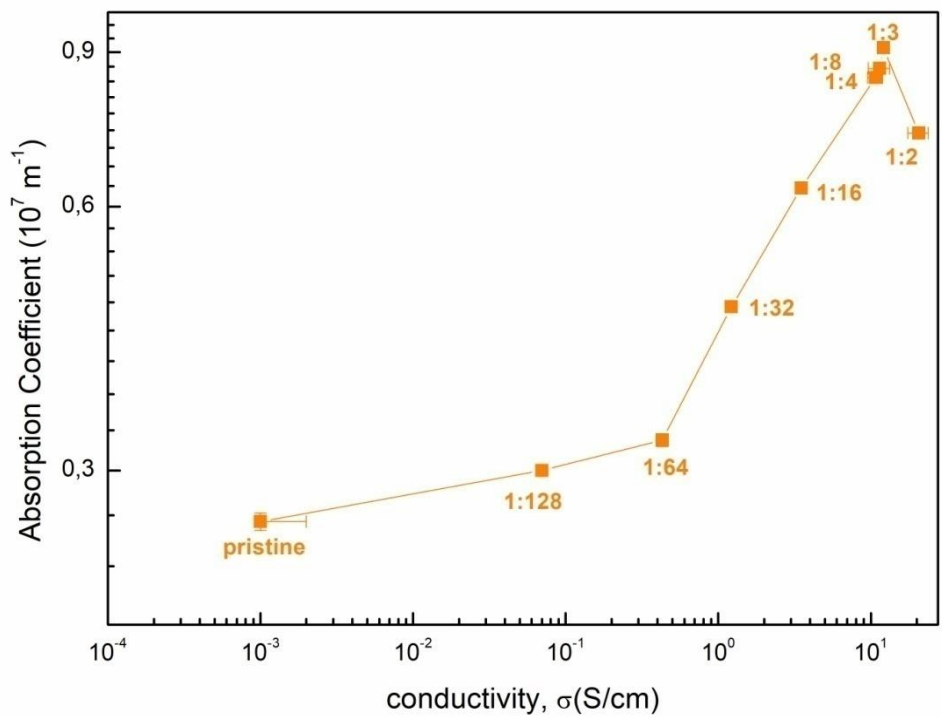

Figure S3. Absorption coefficient versus conductivity below the band-gap energy (700 nm)